\documentclass{article}
\usepackage[utf8]{inputenc}
\usepackage{graphicx}
\usepackage[margin=2cm]{geometry}
\usepackage{caption}
\usepackage{subcaption}
\usepackage{color}
\usepackage{natbib}
\usepackage{amsmath}

\title{Tracking nonlinear solar-wind dynamics over three solar cycles using Wind observations}

\author{
{\small Dario Javier Zamora$^{1,2}$%
\thanks{E-mail: djzamora@conicet.gov.ar} and
Facundo M\'aximo Abaca$^{1,2}$}\\
{\small $^1$ Instituto de F\'isica del Noroeste Argentino,
CONICET and Universidad Nacional de Tucum\'an}\\
{\small $^2$ Departamento de F\'isica, Facultad de Ciencias
Exactas y Tecnolog\'ia, Universidad Nacional de Tucum\'an}\\
{\small Av. Independencia 1800, Tucum\'an, CP 4000, Argentina}
}

\date{\today}

\begin{document}

\maketitle

\begin{abstract}
The solar wind is a turbulent, weakly collisional plasma characterized by non-Gaussian fluctuations, long-range correlations, and multifractal dynamics. We investigate the long-term evolution of these properties using hourly proton density measurements obtained directly by the Wind spacecraft over 1995--2025. The use of a single-spacecraft dataset provides an independent test of previous results derived from the multi-spacecraft OMNI database. The three components of the nonextensive $q$-triplet are estimated within one-year sliding windows shifted monthly: $q_{stat}$ from the distribution of density increments, $q_{rel}$ from the decay of the autocorrelation function, and $q_{sens}$ from the multifractal spectrum. Their mean values, $q_{stat}=1.71\pm0.07$, $q_{rel}=4.62\pm0.52$, and $q_{sens}=-0.45\pm0.27$, confirm the persistent presence of heavy-tailed statistics, slow relaxation, and weakly chaotic multifractal dynamics. The parameters nevertheless exhibit distinct temporal variability and relationships with solar activity. The Fourier spectrum of $q_{stat}$ contains a dominant period of approximately $10.1$ years, while its correlation with the sunspot number is positive and moderate ($r=0.611$). Correlation and mutual-information analyses show that $q_{stat}$ is primarily associated with solar proxies, whereas $q_{sens}$ displays stronger relationships with geomagnetic indices and $q_{rel}$ exhibits weaker dependencies. An anomalous enhancement of $q_{stat}$ around 2004 coincides with a sequence of intense interplanetary disturbances, although possible effects related to Wind's orbital transition must also be considered. These findings demonstrate the robustness of the nonextensive description and show that the statistical and dynamical properties of solar wind proton density show evidence of modulation by solar activity.
\end{abstract}

\section{Introduction}

The solar wind is a rapidly moving stream of charged particles emitted from the solar corona and propagating across the heliosphere. Since the advent of the space age, numerous missions have measured its plasma properties and associated electromagnetic fields. These observations have provided an invaluable basis for advancing our understanding of the solar wind’s physical characteristics and dynamical behavior.

These measurements have revealed that many processes in the solar wind exhibit inherently nonlinear behavior, making it essential to account for the temporal variability of its physical properties. Although conventional approaches are often adequate for stationary or quasi-stationary systems, investigating time-dependent regimes, fluctuations, and scale-invariant structures requires tools from nonlinear dynamics. In this article, we apply nonlinear-dynamics techniques to low-resolution measurements from the Wind mission to investigate the long-term properties of solar-wind plasma over three decades (1995–2025) and their dependence on the solar cycle. This article constitutes a substantially expanded and revised version of the conference proceedings paper presented in \citet{Zamora2026} and includes new results, further methodological developments, and a broader discussion of their physical implications. 

Turbulence is a fundamental property of the solar wind, transferring energy from large injection scales to smaller scales, where it is converted into heat and particle acceleration. This cascade produces power-law spectra and three main nonlinear signatures: heavy-tailed distributions, slow relaxation, and multifractal scaling \citep{Bruno2013}. The inhomogeneous transfer of energy concentrates fluctuations in localized regions, generating heavy-tailed PDFs observed in electron velocity distributions and temperatures, as well as in the heliospheric magnetic-field magnitude \citep{Yoon2024,Stverak2008,Burlaga2004,Burlaga2009,Burlaga2024}. These non-Gaussian states are also associated with slow relaxation processes occurring within the turbulent cascade \citep{Servidio2014,Verscharen2019,Zamora2022}. Moreover, intermittency gives rise to multifractal behavior, which has been detected in magnetic-field fluctuations across different heliocentric distances and solar-cycle phases \citep{Burlaga2003,Burlaga2004b}. Although the solar wind is often treated as weakly compressible, correlations among velocity, temperature, and density suggest that comparable nonlinear features should also occur in proton-density fluctuations, as confirmed across multiple temporal scales \citep{Elliott2016,Borovsky2021,Sorriso-Valvo2017,Zamora2025}.

The paper is structured as follows. Section~2 outlines the foundations of nonextensive statistical mechanics and introduces the $q$-triplet. Section~3 presents the dataset, preprocessing steps, and procedures employed to estimate the $q$-indices. Section~4 discusses the principal results obtained from the proton-density measurements, with emphasis on their statistical properties and variation with solar activity. The conclusions are provided in Section~5.
\section{Theoretical foundations}

Boltzmann--Gibbs (BG) statistical mechanics provides the standard description of systems in thermodynamic equilibrium whose microscopic dynamics are ergodic and dominated by short-range correlations. Under these conditions, fluctuations commonly follow Gaussian statistics and normal diffusion. However, nonlinear nonequilibrium systems may exhibit long-range interactions, memory effects, sensitivity to initial conditions, and strong spatial or temporal correlations, leading to non-Gaussian distributions and anomalous dynamics \citep{Umarov2010}. These properties limit the applicability of the conventional BG framework and motivate the use of generalized statistical approaches.

Nonextensive statistical mechanics has been successfully applied to a wide variety of complex systems \citep{Gell-Mann2004,Tsallis2009,tsallis09a,Vignat2009}, including numerous astrophysical and space-plasma phenomena \citep{plastino1993,Chavanis1998,Scarfone2008,Sahu2012,Rosa2013,Pavlos2018,Zamora2018c,Zamora2020}. In the solar wind, the non-Gaussian distributions observed in magnetic-field increments and other plasma parameters are well described by the $q$-Gaussian distributions predicted by this framework \citep{Burlaga2004,Burlaga2004a,Burlaga2005a,Burlaga2020}. These distributions are closely related to the $\kappa$-distributions traditionally employed in plasma physics, with both formulations becoming mathematically equivalent under an appropriate parameter transformation \citep{Livadiotis2009,Livadiotis2016,Yoon2019,Lazar2021}. Nevertheless, the nonextensive formalism provides a broader theoretical framework that also describes relaxation and multifractal properties.

Within this framework, non-Gaussian statistics, slow relaxation, and sensitivity to initial conditions are characterized by the parameters $(q_{stat},q_{rel},q_{sens})$, collectively known as the $q$-triplet \citep{Tsallis2003,Tsallis2005,Gell-Mann2004}. The statistical component is described by the $q$-Gaussian distribution

\begin{equation}
p(x) \propto
\left[1-(1-q_{stat})\beta x^2\right]^{\frac{1}{1-q_{stat}}},
\label{qgausseq}
\end{equation}

\noindent which acts as the stationary attractor of the correlated dynamics. Values of $q_{stat}>1$ produce heavy-tailed distributions and, in the solar wind, are generally associated with intermittency, coherent structures, shocks, and discontinuities. By contrast, values approaching unity indicate more homogeneous conditions and recover the Gaussian limit.

The relaxation of a macroscopic observable $\Omega$ is described by

\begin{equation}
\Omega(t)=e_{q_{rel}}^{-t/\tau},
\end{equation}

\noindent where

\begin{equation}
e_q^x \equiv
\left[1+(1-q)x\right]^{\frac{1}{1-q}}
\end{equation}

\noindent is the $q$-exponential function. Values of $q_{rel}>1$ represent slower-than-exponential relaxation and indicate that fluctuations remain correlated over extended timescales. In the solar wind, this behavior is consistent with multiscale turbulence and nonlocal interactions, whereas $q_{rel}\rightarrow1$ corresponds to ordinary exponential relaxation and faster decorrelation.

The parameter $q_{sens}$ characterizes the sensitivity of the system to its initial conditions. The separation $\xi(t)$ between initially nearby trajectories evolves according to \citep{Lyra1998}

\begin{equation}
\xi(t)
=
e_{q_{sens}}^{\lambda_q t}
=
\left[1+(1-q_{sens})\lambda_q t\right]^{\frac{1}{1-q_{sens}}},
\end{equation}

\noindent where $\lambda_q$ is the generalized Lyapunov coefficient. For strongly chaotic dynamics with a positive ordinary Lyapunov exponent, $q_{sens}=1$ and the trajectories diverge exponentially. When the ordinary Lyapunov exponent vanishes but $\lambda_q>0$, the system exhibits weak sensitivity characterized by $q_{sens}<1$. Such behavior is commonly related to intermittent and multifractal dynamics, with lower values indicating stronger deviations from homogeneous chaotic behavior.

The BG equilibrium limit is recovered when the triplet becomes $(q_{stat},q_{rel},q_{sens})=(1,1,1)$. The $q$-triplet has been applied to several atmospheric and space-plasma systems, including geomagnetic indices, sunspot and solar-flare time series, magnetospheric dynamics, and solar-wind turbulence \citep{Pavlos2011,Pavlos2012,Karakatsanis2013,Gopinath2018}. Previous analyses of solar-wind magnetic-field measurements reported representative values of $q_{stat}=1.75$, $q_{rel}=4$, and $q_{sens}=-0.5$ \citep{Burlaga2005,Burlaga2013}. More recently, \citet{Zamora2025} identified comparable nonlinear signatures in 17 years of proton-density observations near 1~AU. The present study extends this approach to three decades of Wind measurements (1995--2025) and employs sliding temporal windows to investigate the long-term evolution of the $q$-triplet and its dependence on the solar cycle.

\section{Data analysis}

Unlike our previous analyses \citep{Zamora2025,Zamora2026}, which relied on the OMNI database \citep{King2005}, the present study uses measurements obtained directly from the Wind spacecraft to assess the robustness of the results reported in those works. Although OMNI provides a valuable long-term record, it is a composite dataset assembled from observations acquired by several spacecraft. Its construction therefore involves spacecraft selection, intercalibration, cross-normalization, and other harmonization procedures; depending on the selected data product, propagation time shifts may also be applied to place the measurements at a common reference location. While these procedures are necessary to produce a consistent multi-mission record, they may introduce processing-dependent effects into the probability distributions, temporal correlations, and scaling properties examined here. The use of a single-spacecraft record reduces the influence of cross-mission adjustments and provides a more internally homogeneous dataset for investigating nonlinear behavior over extended periods.

Wind is a spin-stabilized spacecraft launched on November 1, 1994, as part of the Global Geospace Science program within the International Solar Terrestrial Physics initiative \citep{Wilson2021}. Following an initial sequence of geocentric and magnetospheric orbits, the spacecraft was placed in a Lissajous orbit around the L1 Lagrange point in 2004 and subsequently transferred to a halo orbit around L1 in 2020. Its scientific objectives include providing long-term measurements of the plasma, energetic particles, and magnetic field in the near-Earth solar wind, as well as baseline observations near 1~AU for heliospheric studies.

We analyze Wind observations covering the period 1995--2025. Proton plasma parameters are obtained from the Solar Wind Experiment (SWE). The quantity considered here is the proton number density $N_p$, derived through nonlinear fitting of the measured ion current distributions. The original measurements are averaged into one-hour intervals after excluding fill values and measurements identified as invalid by the quality criteria of the corresponding data products. Because Wind followed magnetospheric and geocentric trajectories during the early part of the mission, only intervals classified as valid solar-wind observations are retained.

The three-decade coverage encompasses several phases of solar activity and a broad range of solar-wind conditions. This makes it possible to evaluate the temporal stability of the nonextensive description and determine whether the $q$-triplet exhibits systematic variations across the solar cycle. Although the Wind record is temporally shorter than the multi-spacecraft OMNI dataset employed in our previous work, it offers the advantage of greater instrumental homogeneity and allows the nonlinear properties of proton-density fluctuations to be examined using a consistent set of measurements.

To characterize the temporal evolution of the $q$-triplet, we use overlapping sliding windows rather than dividing the dataset into fixed, non-overlapping intervals. Each window contains one year of observations, and its center is displaced by one month between consecutive estimates. This procedure produces a monthly sampled representation of the annual-scale evolution of the $q$-indices, facilitating the identification of long-term trends, transitions, and possible relationships with solar activity. However, because adjacent windows share eleven months of data, consecutive estimates are not statistically independent and should not be interpreted as independent monthly measurements.

We subsequently compare the temporal evolution of $q_{stat}$, $q_{rel}$, and $q_{sens}$ with solar-activity indicators to investigate how large-scale solar forcing influences non-Gaussianity, long-range temporal correlations, and multifractal behavior in the near-Earth solar wind.

The parameter $q_{stat}$ characterizes the statistical distribution of the fluctuations. For the proton density $N_p$, the consecutive increments are defined as

\begin{equation}
dN_p(i)=N_p(i+1)-N_p(i).
\end{equation}

The increments are normalized using the corresponding local moving average to reduce the influence of slow variations in the background signal. The empirical cumulative distribution function is then fitted with the cumulative form of a $q$-Gaussian distribution. This CDF-based procedure follows the methodology we described in \citet{Abaca2026} and reduces the dependence of the estimate on histogram binning.

The relaxation parameter $q_{rel}$ is determined from the temporal decay of the autocorrelation function. For a time series $N_p(t)$, the normalized autocorrelation coefficient is defined as

\begin{equation}
C(\tau)=
\frac{
\left\langle
[N_p(t+\tau)-\langle N_p\rangle]
[N_p(t)-\langle N_p\rangle]
\right\rangle
}{
\left\langle
[N_p(t)-\langle N_p\rangle]^2
\right\rangle
},
\end{equation}

\noindent where $\tau$ is the time lag and the brackets represent averages over the corresponding annual window. In the nonextensive framework, the correlation function follows a $q$-exponential relaxation and approaches a power law at sufficiently large lags,

\begin{equation}
C(\tau)\propto \tau^\frac{1}{1-q_{rel}},
\end{equation}

Consequently, $q_{rel}$ can be obtained from the slope $s$ of the linear region in a log--log representation.

The parameter $q_{sens}$ is inferred from the multifractal structure of the time series. Statistical moments of the normalized signal are evaluated over a range of temporal scales $\tau$ using moving averages. Their scaling behavior is described by

\begin{equation}
\langle N_p^k\rangle_{\tau}\sim\tau^{s(k)},
\end{equation}

\noindent where $k$ is the moment order and $s(k)$ is the corresponding scaling exponent. A nonlinear dependence of $s(k)$ on $k$ indicates multifractal behavior. The generalized dimensions are obtained from these scaling exponents, and the multifractal spectrum $f(\alpha)$ is subsequently calculated through a Legendre-type transformation, where $\alpha$ is the Hölder exponent.

The limiting singularity exponents, $\alpha_{\min}$ and $\alpha_{\max}$, are used to determine $q_{sens}$ according to

\begin{equation}
\frac{1}{1-q_{sens}}
=
\frac{1}{\alpha_{\min}}
-
\frac{1}{\alpha_{\max}}.
\end{equation}

In practice, the multifractal spectrum is fitted with a suitable concave function and extrapolated to determine its limiting values. The uncertainty in $q_{sens}$ therefore reflects both the statistical quality of the data and the sensitivity of the result to the fitting and extrapolation procedures. Further details of the multifractal method are provided in \citet{Zamora2025} and the references therein.

\section{Results}

\begin{figure}[h!]
\centering
\includegraphics[width=0.65\linewidth]{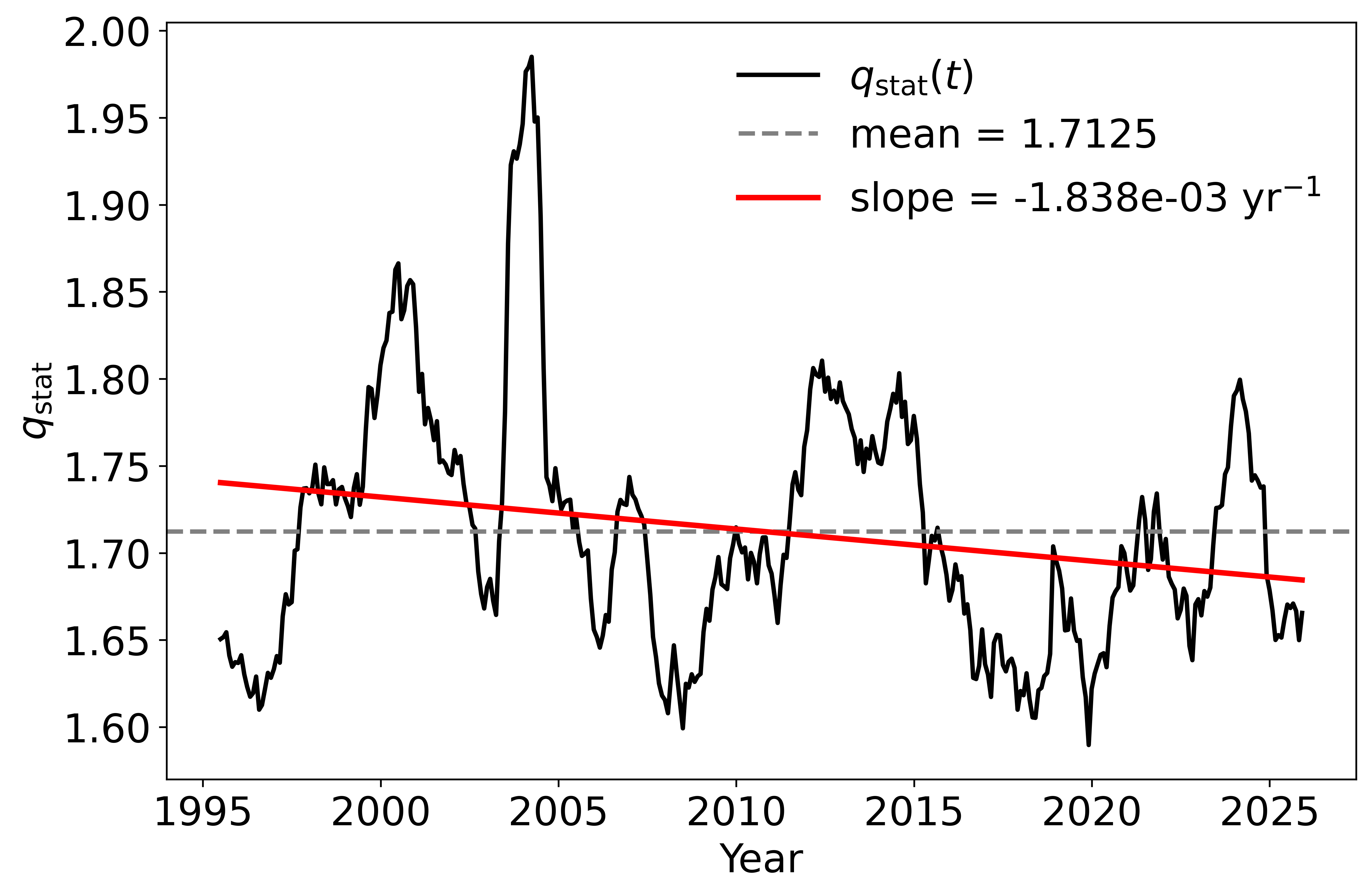}
\caption{Temporal evolution of $q_{stat}$ derived from Wind proton density measurements over 1995--2025 using sliding one-year windows. The dashed gray line indicates the mean value, while the red line represents the linear trend.}
\label{timeserieqstat}
\end{figure}

The temporal evolution of $q_{stat}$ derived from Wind proton density measurements over 1995--2025 is shown in Fig.~\ref{timeserieqstat}. Despite substantial variability, $q_{stat}$ remains above unity throughout the interval, confirming the persistent non-Gaussian and heavy-tailed character of the fluctuations. The series also exhibits a weak decreasing trend, suggesting a gradual modulation of the solar wind statistical properties. Since values closer to unity correspond to lighter distribution tails, this decrease may indicate a modest reduction in the occurrence of extreme fluctuations. Although longer-term solar modulations, such as the Gleissberg cycle, could contribute to this behavior \citep{Gleissberg1939}, the 30-year Wind record is insufficient to establish such a connection conclusively.

\begin{figure}[h!]
\centering
\includegraphics[width=0.65\linewidth]{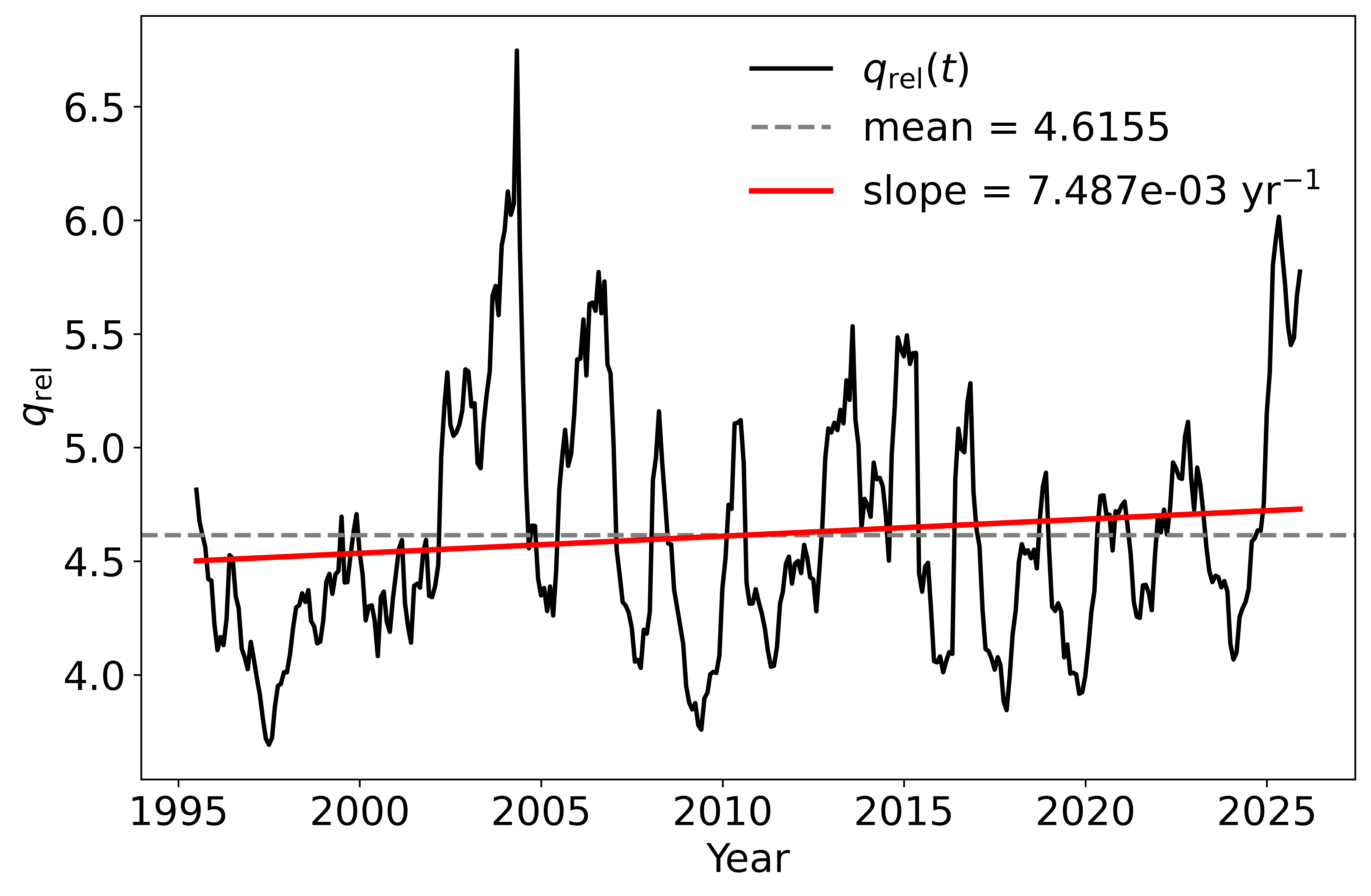}
\caption{Temporal evolution of $q_{rel}$ derived from Wind proton density measurements over 1995--2025 using sliding one-year windows. The dashed gray line indicates the mean value, while the red line represents the linear trend.}
\label{timeserieqrel}
\end{figure}

\begin{figure}[h!]
\centering
\includegraphics[width=0.65\linewidth]{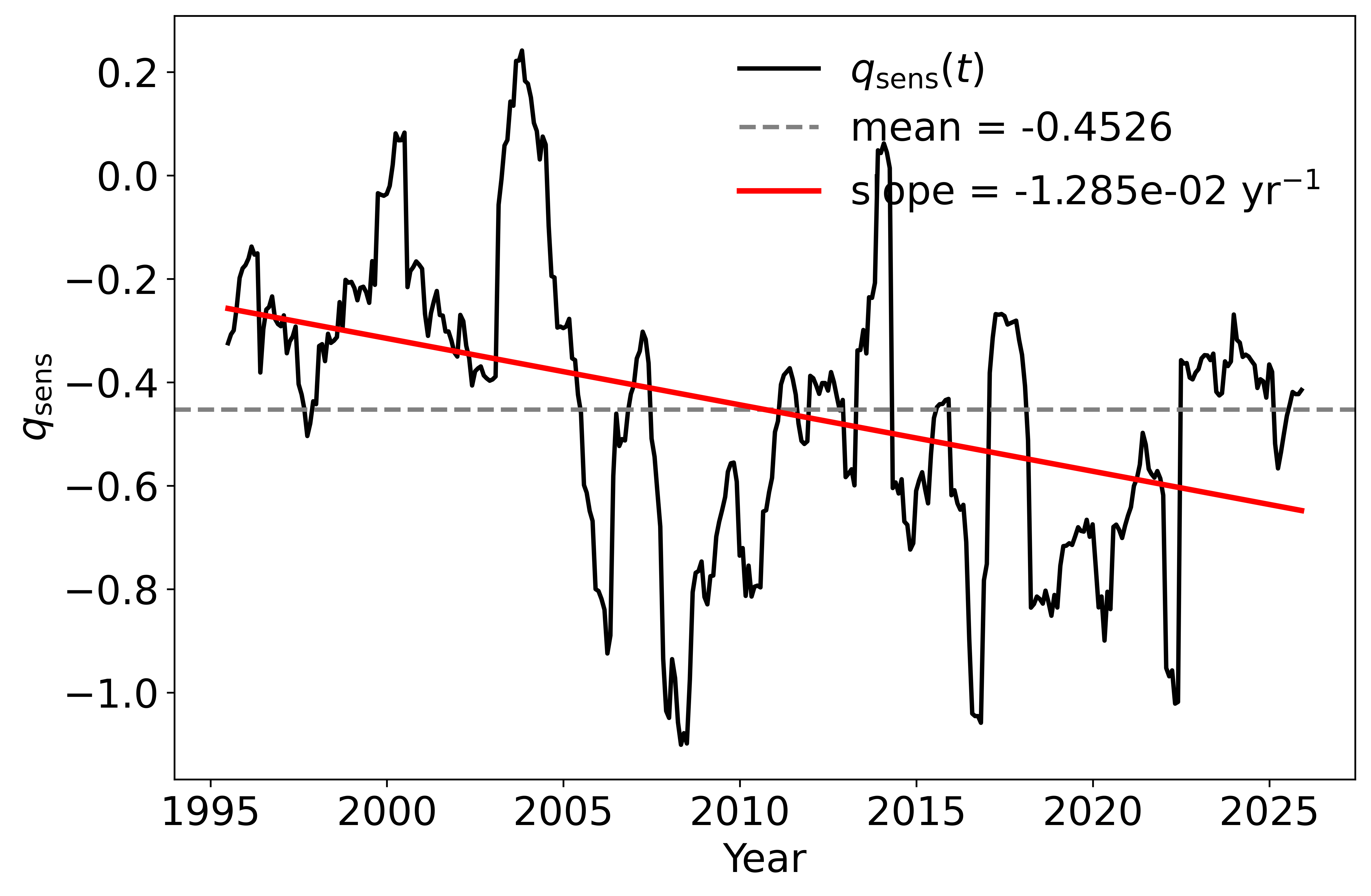}
\caption{Temporal evolution of $q_{sens}$ derived from Wind proton density measurements over 1995--2025 using sliding one-year windows. The dashed gray line indicates the mean value, while the red line represents the linear trend.}
\label{timeserieqsens}
\end{figure}

The temporal evolution of $q_{rel}$ and $q_{sens}$ derived from Wind proton density measurements is shown in Figs.~\ref{timeserieqrel} and \ref{timeserieqsens}, respectively. The values of $q_{rel}$ remain well above unity and exhibit a positive trend, indicating persistent long-range correlations and a gradual slowing of the relaxation dynamics. Conversely, $q_{sens}$ remains below unity and is predominantly negative, and shows a weak decreasing trend, consistent with weak chaos and multifractal dynamics. This behavior may reflect an increasing persistence and scale-dependent organization of solar wind fluctuations.

\begin{figure}[h!]
\centering
\includegraphics[width=0.65\linewidth]{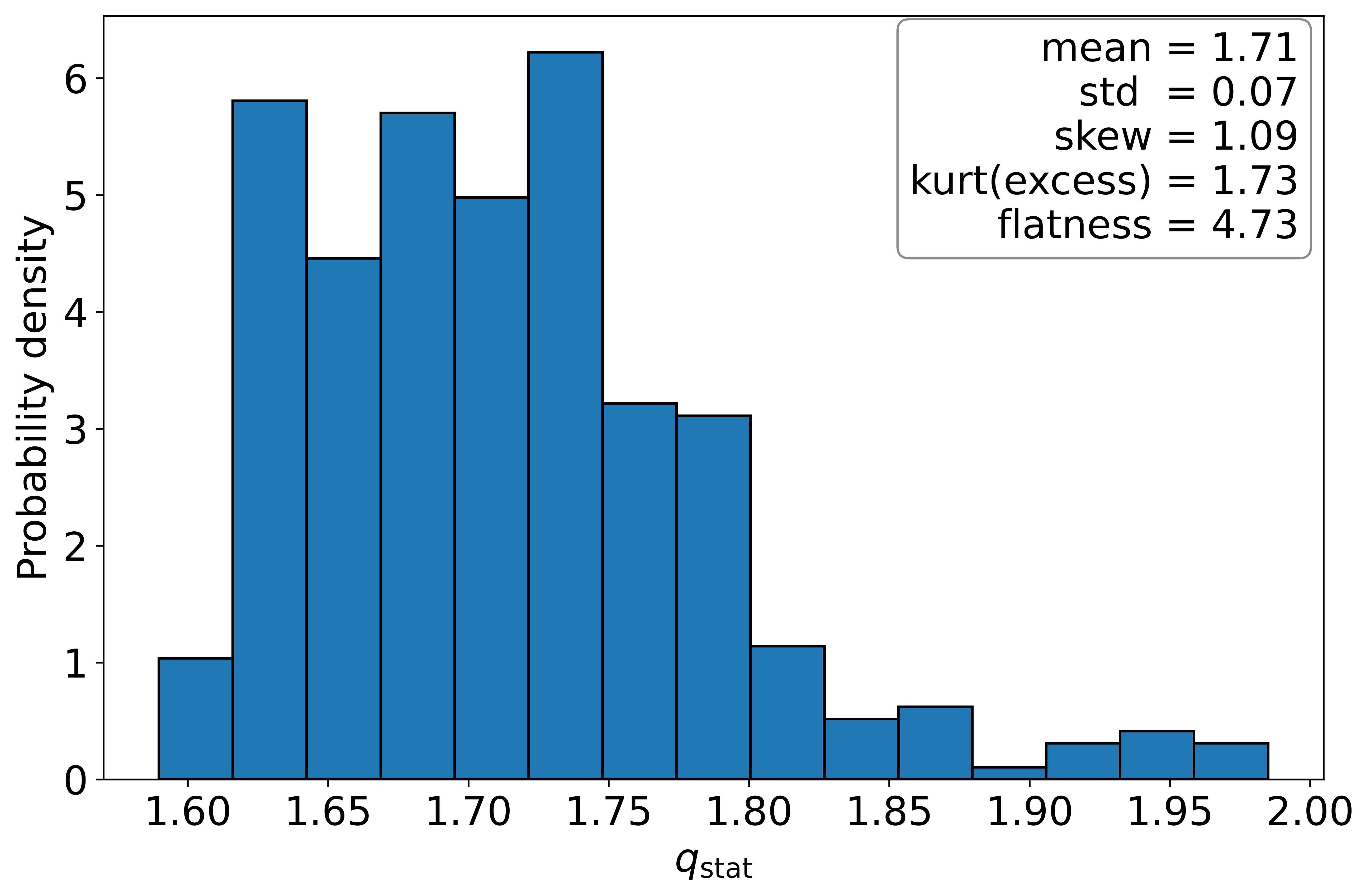}
\caption{Probability density function of $q_{stat}$ derived from Wind proton density measurements over 1995--2025.}
\label{PDFqstat}
\end{figure}

\begin{figure}[h!]
\centering
\includegraphics[width=0.65\linewidth]{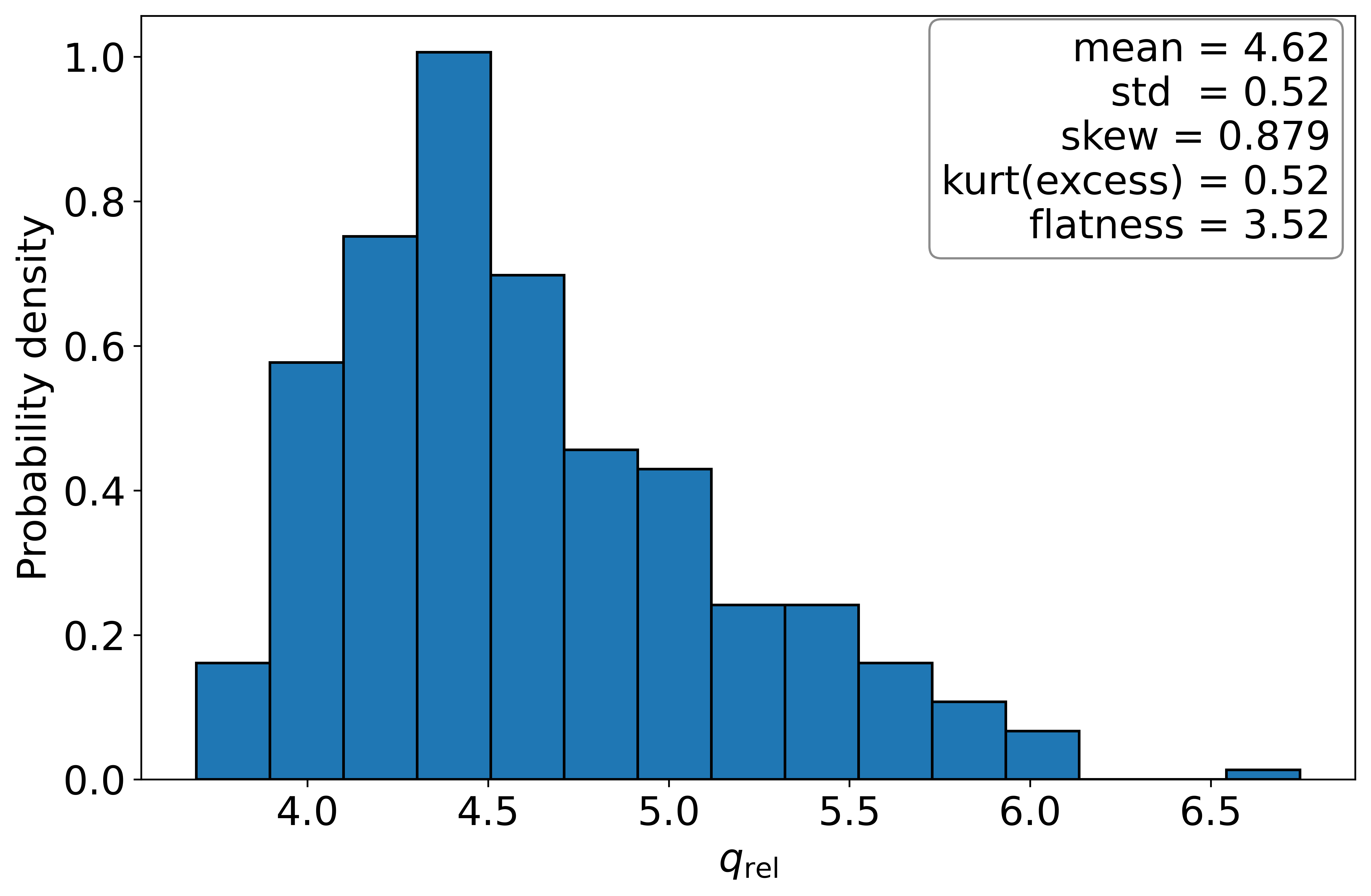}
\caption{Probability density function of $q_{rel}$ derived from Wind proton density measurements over 1995--2025.}
\label{PDFqrel}
\end{figure}

\begin{figure}[h!]
\centering
\includegraphics[width=0.65\linewidth]{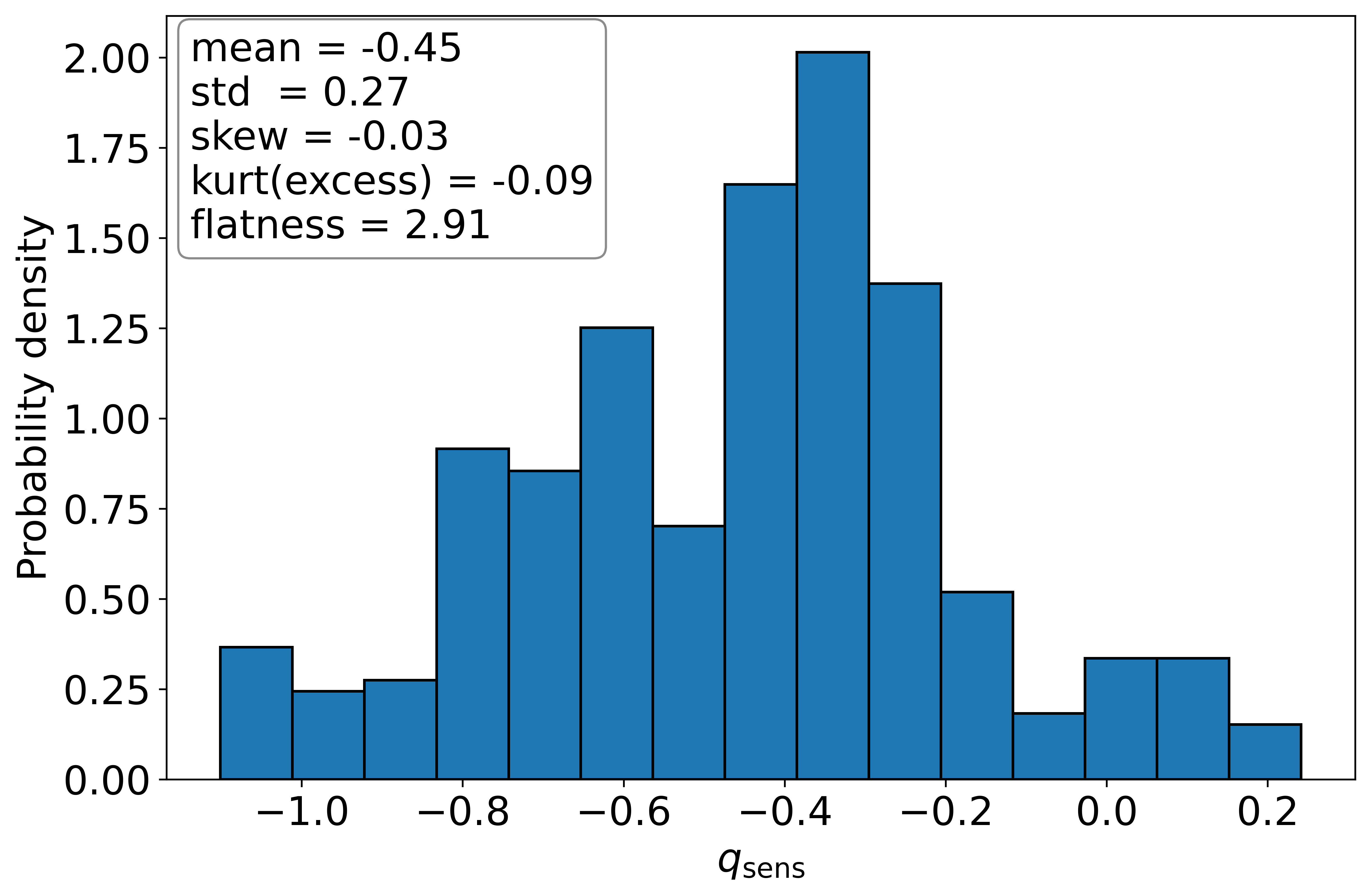}
\caption{Probability density function of $q_{sens}$ derived from Wind proton density measurements over 1995--2025.}
\label{PDFqsens}
\end{figure}

Taken together, these results show that the three components of the $q$-triplet exhibit distinct long-term statistical behavior. Their probability density functions are presented in Figs.~\ref{PDFqstat}--\ref{PDFqsens}. The distribution of $q_{stat}$ is relatively narrow, with a mean of $1.71$ and a standard deviation of $0.07$. Its positive skewness ($1.09$) and flatness ($4.73$) reveal a pronounced right tail and leptokurtic behavior, indicating occasional intervals of enhanced non-Gaussianity (around year 2004 as can be seen in Fig. \ref{timeserieqstat}). The $q_{rel}$ distribution is broader, with a mean of $4.62$ and a standard deviation of $0.52$, and is also positively skewed ($0.879$) and leptokurtic ($F=3.52$). Its extended right tail reflects sporadic intervals characterized by particularly slow relaxation and stronger temporal correlations particularly around 2004 (Fig. \ref{timeserieqrel}). In contrast, $q_{sens}$ has a mean of $-0.45$ and a standard deviation of $0.27$, with negligible skewness ($-0.03$) and a flatness close to the Gaussian value (remember that a normal distribution has flatness coefficient $F=3$). Thus, $q_{sens}$ displays an approximately symmetric and nearly normal distribution. Overall, the parameters remain statistically well constrained, although $q_{stat}$ and $q_{rel}$ exhibit occasional excursions toward high values. Since consecutive sliding windows overlap, these distributions should be interpreted as descriptive statistics of the temporal evolution rather than as distributions of fully independent estimates.

\begin{figure}[h!]
\centering
\includegraphics[width=0.65\linewidth]{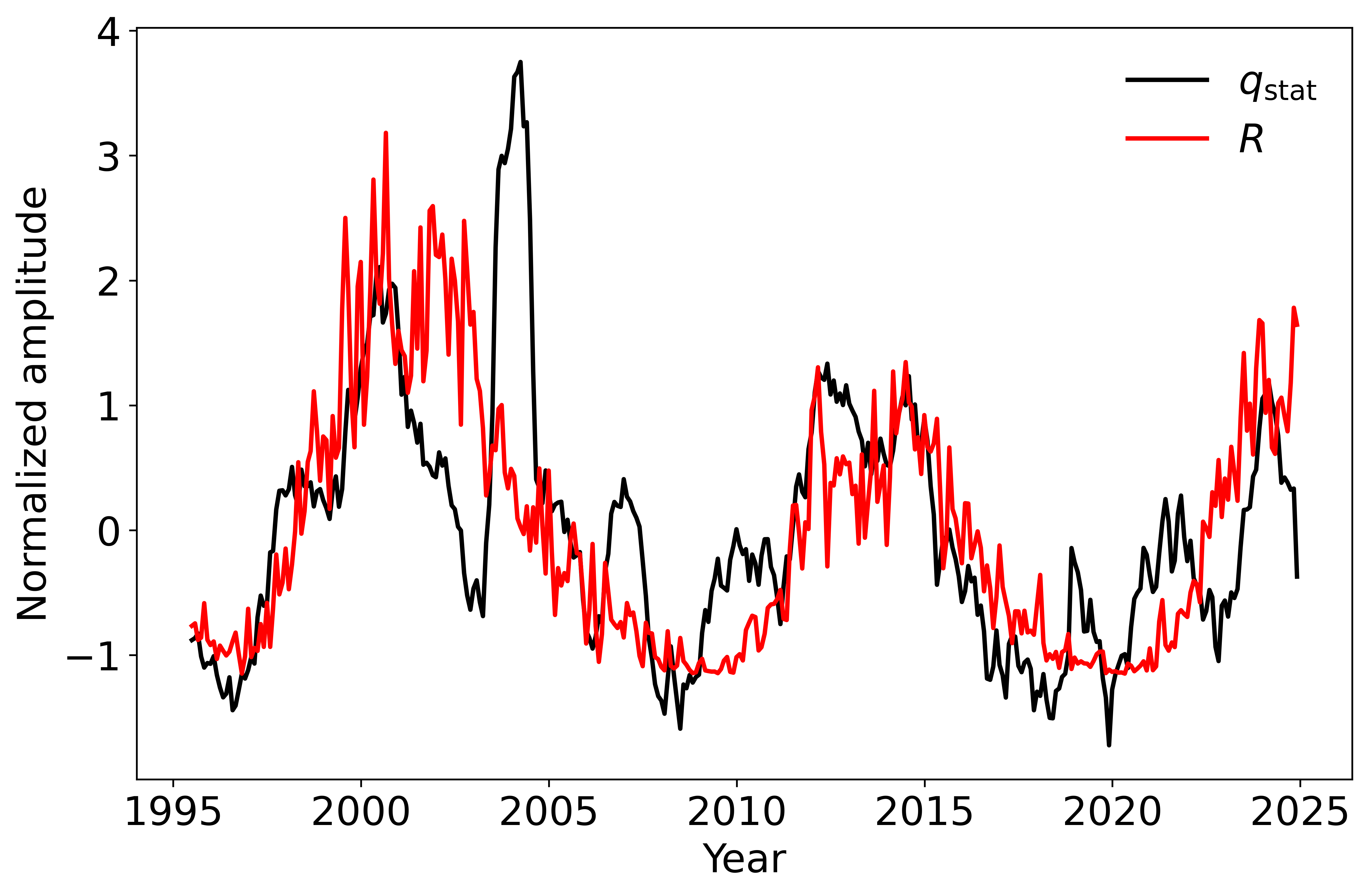}
\caption{Normalized time series of $q_{stat}$ derived from Wind proton density measurements (black) and sunspot number $R$ (red) over 1995--2025. Their broad oscillatory behavior suggests a modulation associated with the solar cycle, despite differences in phase and amplitude.}
\label{R}
\end{figure}

Figure~\ref{R} compares the normalized time series of $q_{stat}$ and the sunspot number $R$ over 1995--2025. Both signals exhibit broad variations compatible with the $\sim$11-year solar cycle, with similar minima around 1996, 2009, and 2020 and enhanced values during periods of greater solar activity. Nevertheless, the agreement is not exact: phase shifts, amplitude differences, and isolated features—particularly the pronounced $q_{stat}$ peak around 2004—are also observed. The overall correspondence suggests that solar activity modulates the non-Gaussian properties of proton density fluctuations, although it is not their only controlling factor.

The pronounced enhancement in $q_{stat}$ around 2004 may be associated with the clustering of several extreme solar wind disturbances within the corresponding one-year sliding windows. In particular, these windows include the intense interplanetary shocks and ICMEs associated with the Halloween storms of late 2003 \citep{Skoug2004}, as well as major CME sequences observed during July and November 2004 \citep{Ngwira2012,Tsurutani2008}. The resulting shocks, compression regions, and abrupt proton density variations may enhance the tails of the increment distributions and consequently increase $q_{stat}$. Nevertheless, Wind completed its insertion into an L1 orbit during this period \citep{Wilson2021}; therefore, possible changes in data coverage and sampling must also be examined before assigning the enhancement exclusively to solar activity.

\begin{figure}[h!]
\centering
\includegraphics[width=0.65\linewidth]{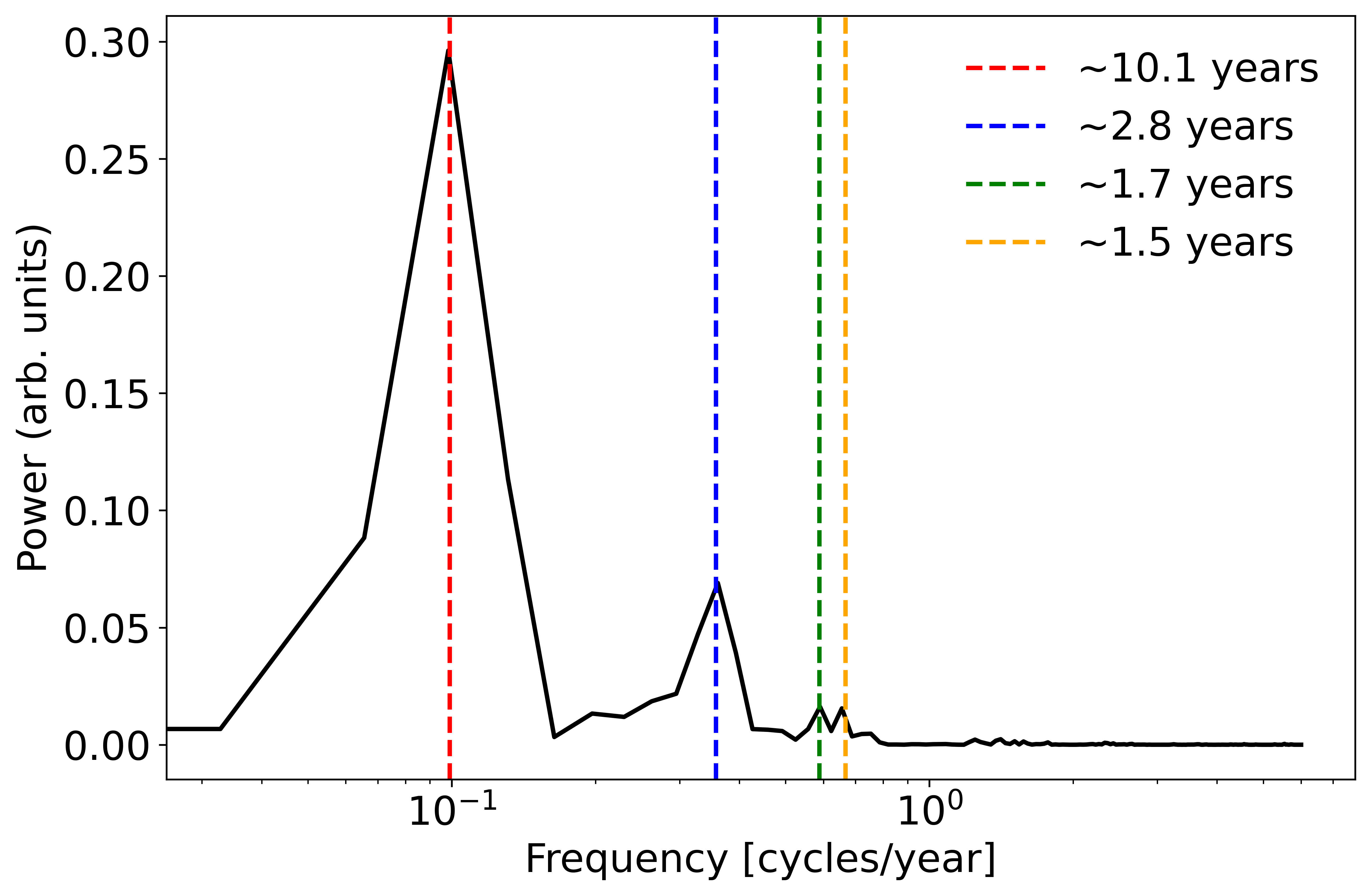}
\caption{Fourier power spectrum of the $q_{stat}$ time series derived from Wind proton density measurements. The dashed vertical lines indicate the dominant $\sim10.1$-year periodicity and the weaker secondary periods at approximately $2.8$, $1.7$, and $1.5$ years.}
\label{Fourierqstat}
\end{figure}

\begin{figure}[h!]
\centering
\includegraphics[width=0.65\linewidth]{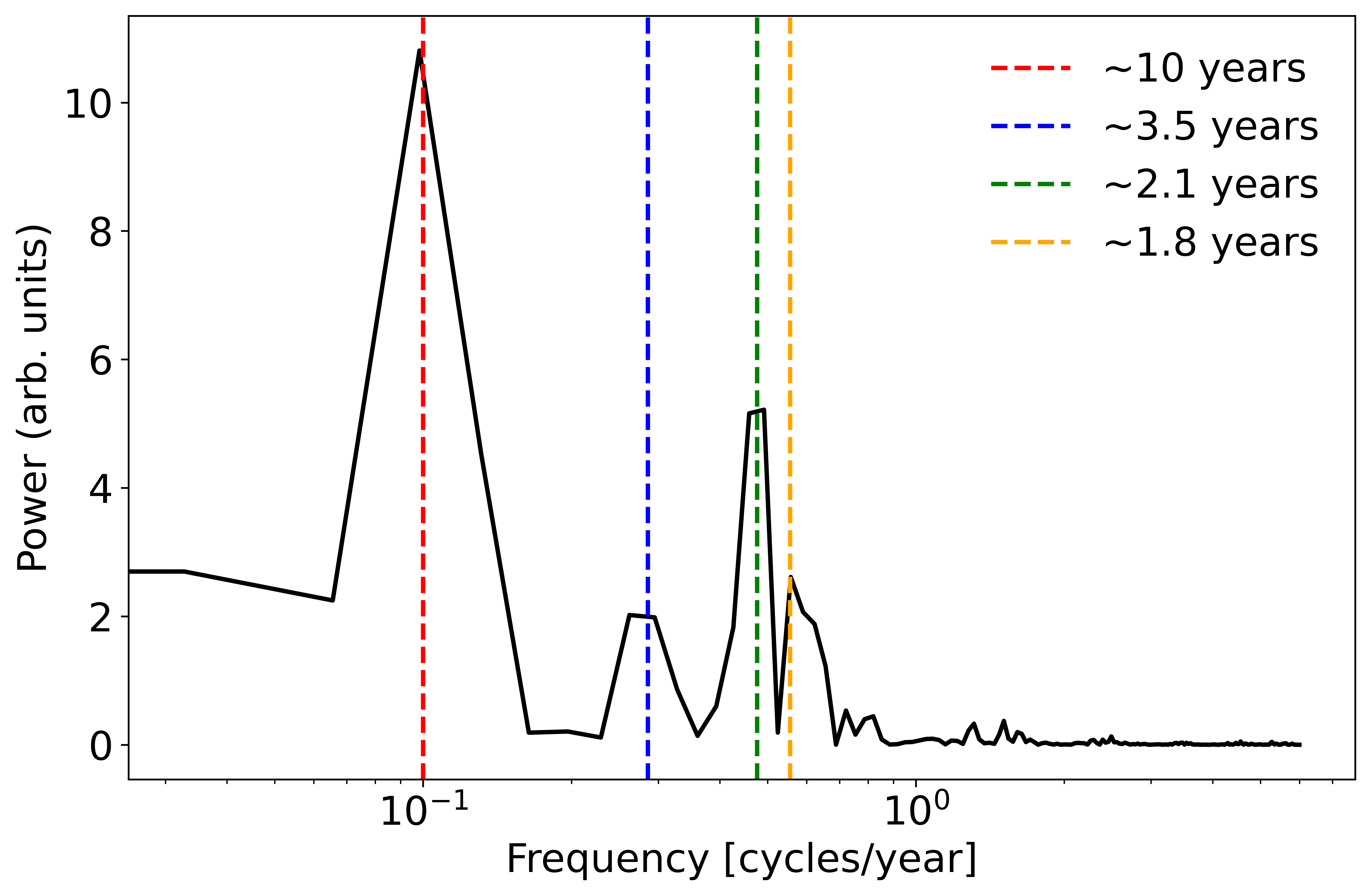}
\caption{Fourier power spectrum of the $q_{rel}$ time series derived from Wind proton density measurements. The dashed vertical lines indicate the dominant $\sim10$-year periodicity and the weaker secondary periods.}
\label{Fourierqrel}
\end{figure}

\begin{figure}[h!]
\centering
\includegraphics[width=0.65\linewidth]{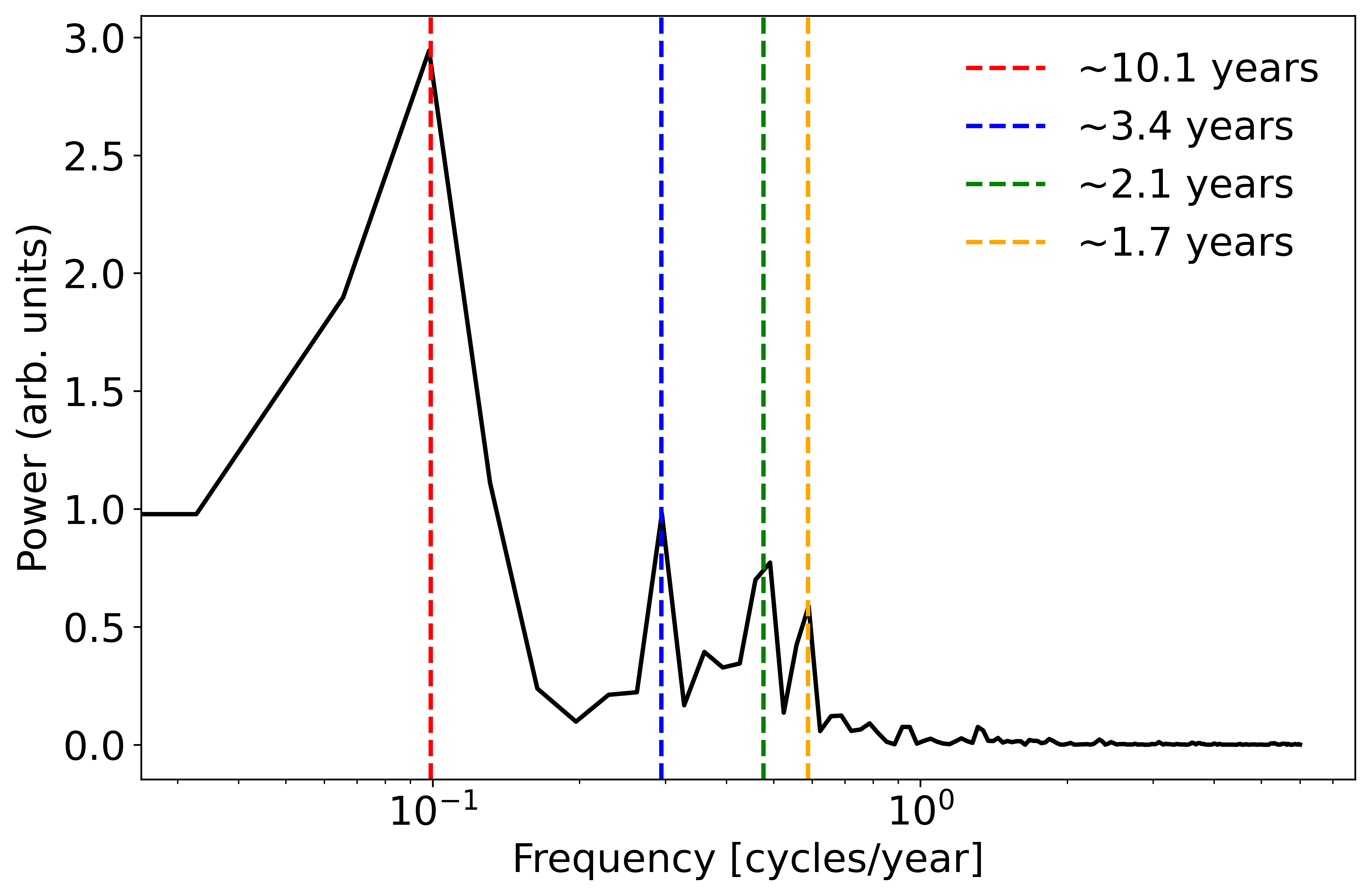}
\caption{Fourier power spectrum of the $q_{sens}$ time series derived from Wind proton density measurements. The dashed vertical lines indicate the dominant $\sim10.1$-year periodicity and the weaker secondary periods.}
\label{Fourierqsens}
\end{figure}

The qualitative correspondence between $q_{stat}$ and solar activity is supported by its Fourier power spectrum, shown in Fig.~\ref{Fourierqstat}. The dominant peak occurs at approximately $0.099$ cycles/year, corresponding to a period of $\sim10.1$ years and therefore consistent with solar-cycle modulation. Much weaker peaks are found at periods of approximately $2.8$, $1.7$, and $1.5$ years, which may be related to previously reported mid-term heliospheric variability \citep{Richardson1994,Valdes-Galicia2008}. Similar decadal peaks are obtained for $q_{rel}$ and $q_{sens}$, as shown in Figs. \ref{Fourierqrel} and \ref{Fourierqsens}.

These results should be treated with caution because the 30-year Wind record contains only about three solar cycles. Moreover, the overlapping one-year windows introduce temporal smoothing and suppress periodicities shorter than one year. Thus, Fourier analysis supports—but does not by itself establish—the dependence of the $q$-triplet on solar activity; correlation and coherence analyses provide complementary tests of this relationship.

\begin{figure}[h!]
\centering
\includegraphics[width=0.65\linewidth]{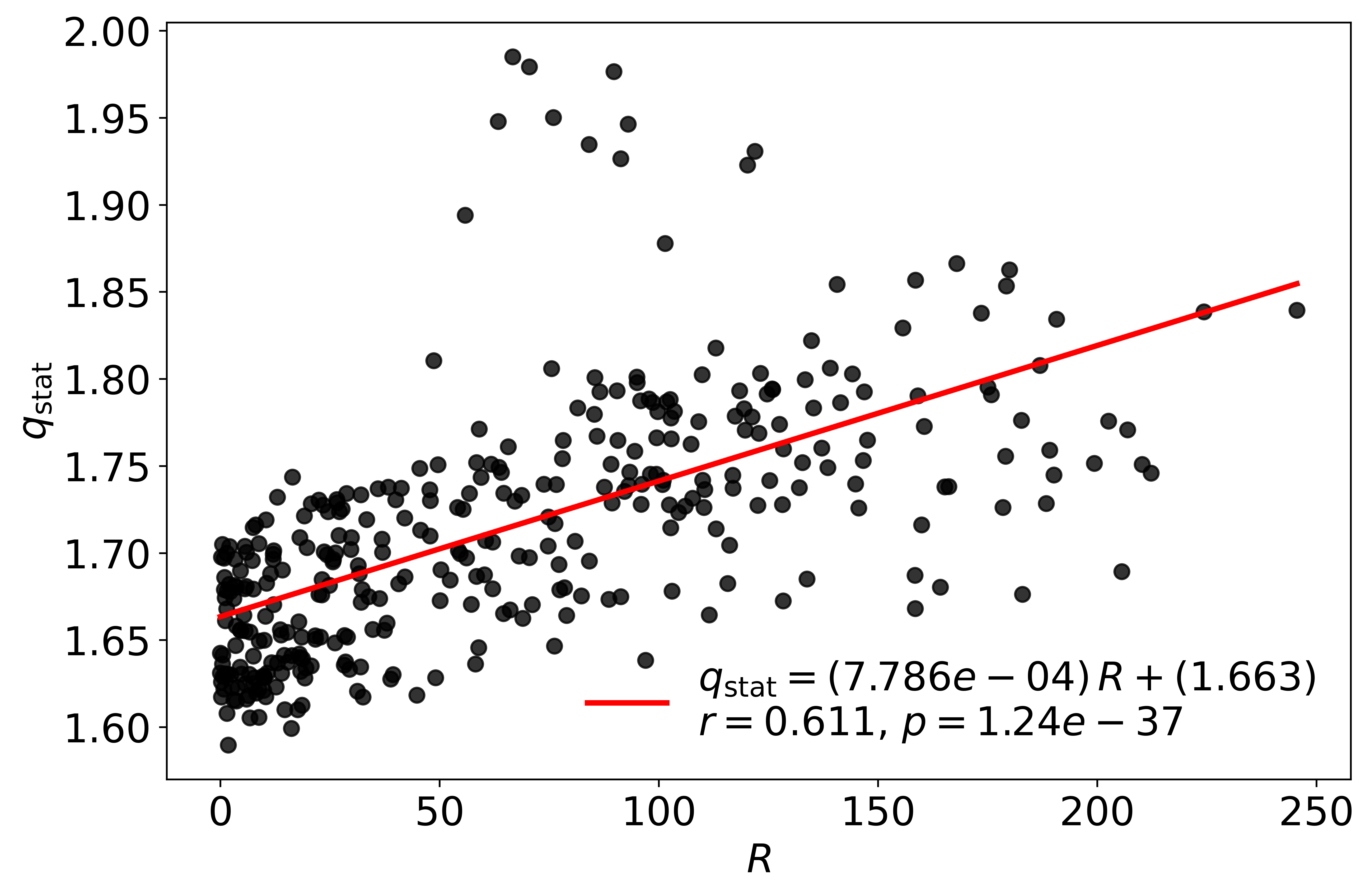}
\caption{Scatter plot of $q_{stat}$ as a function of the sunspot number $R$. The red line represents the linear fit.}
\label{scatter}
\end{figure}

Figure~\ref{scatter} shows a moderate positive relationship between $q_{stat}$ and the sunspot number $R$. The linear regression yields a slope of $7.786\times10^{-4}$ and a Pearson coefficient $r=0.611$ ($p=1.24\times10^{-37}$), indicating that larger values of $q_{stat}$ are generally associated with higher solar activity. The isolated points above the main distribution correspond to the overlapping windows around 2004 and are associated with the exceptional CME and interplanetary shock activity discussed above \citep{Skoug2004,Ngwira2012,Tsurutani2008}. These departures show that $q_{stat}$ is influenced not only by the overall solar-cycle level but also by individual transient disturbances.

\begin{figure*}[h!]
\centering
\includegraphics[width=\linewidth]{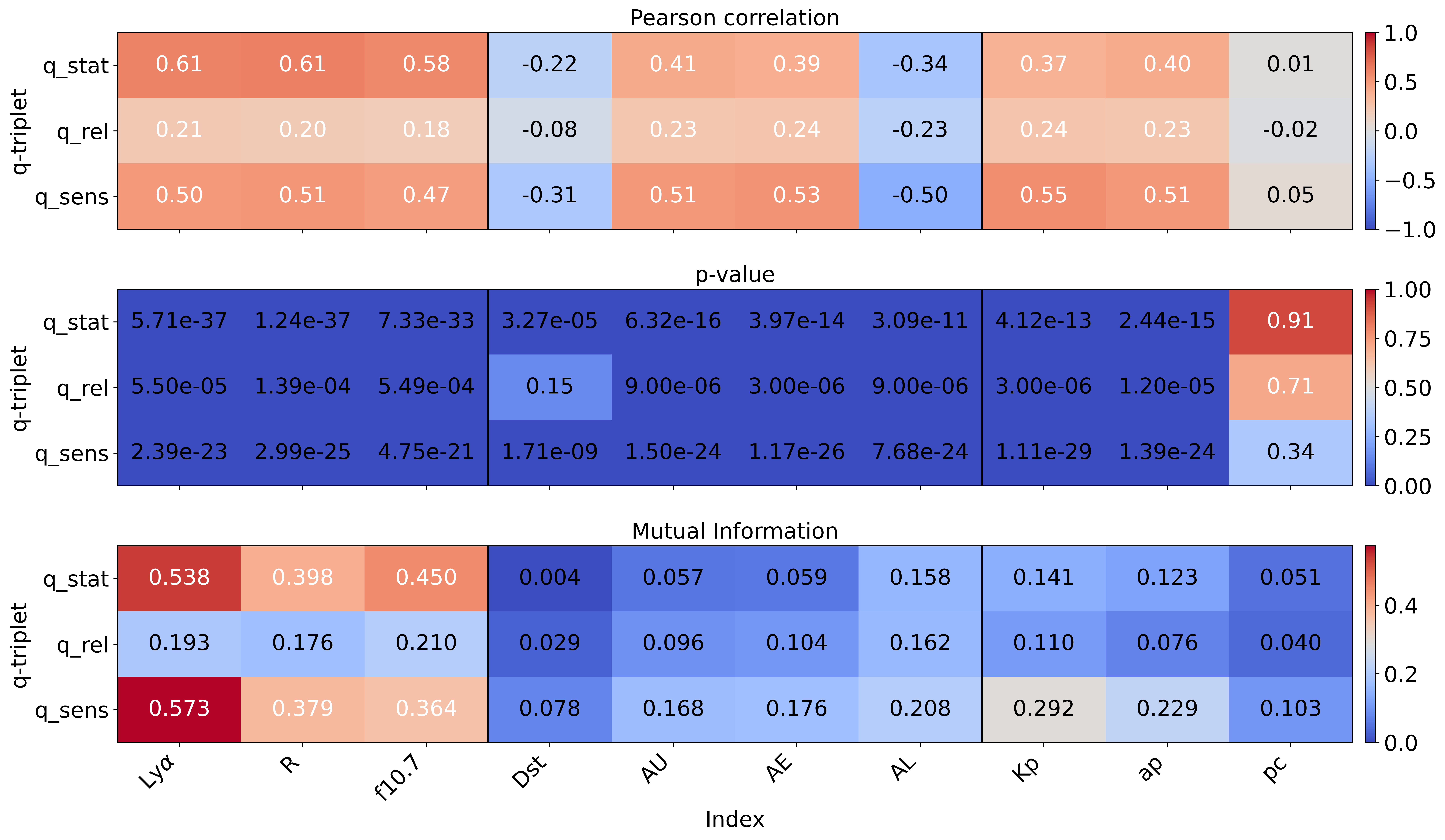}
\caption{Correlation analysis between the $q$-triplet parameters derived from Wind proton density measurements and selected solar and geomagnetic indices. The panels show the Pearson correlation coefficient, its nominal $p$-value, and the mutual information.}
\label{Correlogram}
\end{figure*}

To quantify the dependence of the $q$-triplet on solar and geomagnetic activity, Fig.~\ref{Correlogram} presents the Pearson correlation coefficient, its nominal $p$-value, and the mutual information for several activity indices. The latter captures both linear and nonlinear statistical dependencies, while the Pearson coefficient captures only linear correlations.

The results reveal distinct behaviors among the three parameters. The strongest correlations of $q_{stat}$ occur with the solar proxies Ly$\alpha$, $R$, and $F10.7$, which also display the largest mutual information values. Conversely, $q_{sens}$ is moderately correlated with both solar and geomagnetic activity, reaching $r=0.55$ for Kp and $r=0.53$ for AE. Its negative correlations with Dst and AL reflect the sign convention of these indices, whose increasingly negative values indicate stronger geomagnetic disturbances. The relationships involving $q_{rel}$ are considerably weaker, while pc shows no significant linear correlation with any component of the triplet. Overall, the mutual information analysis supports the coupling of $q_{stat}$ primarily with solar activity and of $q_{sens}$ with both solar and magnetospheric dynamics.

\section{Conclusions}

In this work, we investigated the long-term evolution of the nonextensive $q$-triplet derived from Wind proton density measurements over 1995--2025. Unlike our previous studies based on the multi-spacecraft OMNI database \citep{Zamora2025,Zamora2026}, the present analysis uses measurements obtained directly from a single spacecraft, providing an independent test of the robustness of the nonextensive characterization. Sliding one-year windows shifted monthly were employed to track the temporal variability of $q_{stat}$, $q_{rel}$, and $q_{sens}$ across approximately three solar cycles.

The mean values obtained, $q_{stat}=1.71\pm0.07$, $q_{rel}=4.62\pm0.52$, and $q_{sens}=-0.45\pm0.27$, are consistent with the $q$-triplet values previously reported by Burlaga and collaborators for magnetic-field fluctuations in the distant solar wind and heliosheath \citep{Burlaga2005,Burlaga2013}. The persistent conditions $q_{stat}>1$, $q_{rel}>1$, and $q_{sens}<1$ indicate, respectively, non-Gaussian heavy-tailed fluctuations, slow relaxation with long-range temporal correlations, and weakly chaotic multifractal dynamics. Their different temporal trends and probability distributions further show that the three indices describe complementary aspects of the solar wind rather than a single underlying property.

The results also provide evidence that the $q$-triplet is modulated by solar and geomagnetic activity. The Fourier spectrum of $q_{stat}$ exhibits a dominant period of approximately $10.1$ years, consistent with the solar cycle, while its correlation with the sunspot number is moderate and positive ($r=0.611$). More generally, $q_{stat}$ is most strongly associated with solar proxies such as Ly$\alpha$, $R$, and $F10.7$, whereas $q_{sens}$ displays stronger relationships with geomagnetic indices, particularly Kp, AE, AU, ap, and AL. In contrast, $q_{rel}$ shows weaker and less systematic dependencies. Mutual information supports the existence of additional nonlinear coupling beyond the linear correlations.

The isolated high values of $q_{stat}$ and $q_{rel}$ around 2004 may be related to the intense shocks and CME sequences observed from late 2003 through 2004 \citep{Skoug2004,Ngwira2012,Tsurutani2008}, whose abrupt density variations enhance the tails of the increment distributions. Nevertheless, because this interval also coincides with Wind's final insertion into its L1 orbit, possible effects associated with data coverage and sampling should be examined before attributing the enhancement exclusively to solar activity.

Overall, the agreement with previous OMNI-based results supports the robustness of the nonextensive properties of the solar wind against the choice of database. However, the limited temporal coverage, the strong overlap between consecutive windows, and the smoothing imposed by their one-year duration require caution when assessing statistical significance and short-period variability. Future work should incorporate lagged correlation, wavelet coherence, surrogate-data tests, and analyses using alternative window lengths, as well as extend the same framework to other solar wind variables. Such studies may clarify how solar forcing, transient structures, and turbulent dynamics jointly determine the long-term evolution of the $q$-triplet.

\section*{Data availability}
The proton-density measurements analyzed in this study were obtained from the \textit{Wind Solar Wind Experiment (SWE) 92-sec Definitive Solar Wind Proton Data} product, provided by NASA’s Space Physics Data Facility (SPDF). After removing fill values and observations flagged as invalid, the original measurements were averaged to a temporal resolution of one hour. The dataset is publicly available at doi:10.48322/nasd-j276. Codes, results and analyses presented in this paper are available upon request.

\section*{Acknowledgment}
We thank Jaroslav Urb\'ar for providing access to the data. This work was financially supported by CONICET (Argentina).

\bibliographystyle{apalike}
\bibliography{solarwind_clean}

\end{document}